\documentclass[11pt]{article}\textwidth 6.5in\textheight 9in
\usepackage{amssymb}\usepackage[colorlinks]{hyperref}\usepackage{color}
	\usepackage{stmaryrd}\usepackage{mathrsfs}
		\usepackage{graphicx}
	\usepackage{amsmath}
\usepackage{caption}
    \usepackage{subcaption}
    \usepackage{listings}
 
\begin{document}
\setlength{\captionmargin}{27pt}
\newcommand\hreff[1]{\href {http://#1} {\small http://#1}}
\newcommand\trm[1]{{\bf\em #1}} \newcommand\emm[1]{{\ensuremath{#1}}}
\newcommand\prf{\paragraph{Proof.}}\newcommand\qed{\hfill\emm\blacksquare}

\newtheorem{thr}{Theorem} 
\newtheorem{lmm}{Lemma}
\newtheorem{cor}{Corollary}
\newtheorem{con}{Conjecture} 
\newtheorem{prp}{Proposition}

\newtheorem{blk}{Block}
\newtheorem{dff}{Definition}
\newtheorem{asm}{Assumption}
\newtheorem{rmk}{Remark}
\newtheorem{clm}{Claim}
\newtheorem{exm}{Example}

\newcommand\Ks{\mathbf{Ks}} 
\newcommand{\ab}{a\!b}
\newcommand{\yx}{y\!x}
\newcommand{\yux}{y\!\underline{x}}

\newcommand\floor[1]{{\lfloor#1\rfloor}}\newcommand\ceil[1]{{\lceil#1\rceil}}

\newcommand{\lea}{<^+}
\newcommand{\gea}{>^+}
\newcommand{\eqa}{=^+}

\newcommand{\lel}{<^{\log}}
\newcommand{\gel}{>^{\log}}
\newcommand{\eql}{=^{\log}}

\newcommand{\lem}{\stackrel{\ast}{<}}
\newcommand{\gem}{\stackrel{\ast}{>}}
\newcommand{\eqm}{\stackrel{\ast}{=}}

\newcommand\edf{{\,\stackrel{\mbox{\tiny def}}=\,}}
\newcommand\edl{{\,\stackrel{\mbox{\tiny def}}\leq\,}}
\newcommand\then{\Rightarrow}

\newcommand\C{\mathbf{C}} 

\renewcommand\chi{\mathcal{H}}
\newcommand\km{{\mathbf {km}}}\renewcommand\t{{\mathbf {t}}}
\newcommand\KM{{\mathbf {KM}}}\newcommand\m{{\mathbf {m}}}
\newcommand\md{{\mathbf {m}_{\mathbf{d}}}}\newcommand\mT{{\mathbf {m}_{\mathbf{T}}}}
\newcommand\K{{\mathbf K}} \newcommand\I{{\mathbf I}}

\newcommand\II{\hat{\mathbf I}}
\newcommand\Kd{{\mathbf{Kd}}} \newcommand\KT{{\mathbf{KT}}} 
\renewcommand\d{{\mathbf d}} 
\newcommand\D{{\mathbf D}}

\newcommand\w{{\mathbf w}}
\newcommand\Cs{\mathbf{Cs}} \newcommand\q{{\mathbf q}}
\newcommand\E{{\mathbf E}} \newcommand\St{{\mathbf S}}
\newcommand\M{{\mathbf M}}\newcommand\Q{{\mathbf Q}}
\newcommand\ch{{\mathcal H}} \renewcommand\l{\tau}
\newcommand\tb{{\mathbf t}} \renewcommand\L{{\mathbf L}}
\newcommand\bb{{\mathbf {bb}}}\newcommand\Km{{\mathbf {Km}}}
\renewcommand\q{{\mathbf q}}\newcommand\J{{\mathbf J}}
\newcommand\z{\mathbf{z}}

\newcommand\B{\mathbf{bb}}\newcommand\f{\mathbf{f}}
\newcommand\hd{\mathbf{0'}} \newcommand\T{{\mathbf T}}
\newcommand\R{\mathbb{R}}\renewcommand\Q{\mathbb{Q}}
\newcommand\N{\mathbb{N}}\newcommand\BT{\{0,1\}}
\newcommand\FS{\BT^*}\newcommand\IS{\BT^\infty}
\newcommand\FIS{\BT^{*\infty}}
\renewcommand\S{\mathcal{C}}\newcommand\ST{\mathcal{S}}
\newcommand\UM{\nu_0}\newcommand\EN{\mathcal{W}}

\newcommand{\supp}{\mathrm{Supp}}

\newcommand\lenum{\lbrack\!\lbrack}
\newcommand\renum{\rbrack\!\rbrack}

\newcommand\h{\mathbf{h}}
\renewcommand\qed{\hfill\emm\square}
\renewcommand\i{\mathbf{i}}
\newcommand\p{\mathbf{p}}
\renewcommand\q{\mathbf{q}}
\title{\vspace*{-3pc} 22 Examples of Solution Compression via Derandomization}

\author {Samuel Epstein\footnote{JP Theory Group. samepst@jptheorygroup.org}}

\maketitle
\begin{abstract}
We provide bounds on the compression size of the solutions to 22 problems in computer science. For each problem, we show that solutions exist with high probability, for some simple probability measure. Once this is proven, derandomization can be used to prove the existence of a simple solution.
\end{abstract}
\section{Introduction}
This paper showcases different bounds on the compression level of problems. An example problem is {\sc k-Sat}, which is a conjunction of $m$ clauses, each consisting of exactly $k$ literals. A literal is either a variable $v_i$ or the negation of a variable $\overline{v}_i$. Say there are $n$ variables. The goal is to find an assignment of $n$ variables that satisfy the conjunction. This is an NP-complete problem.

One of the goals in the  field of algorithmic information theory, is to prove upper bounds on the smallest compression size of objects with combinatorial properties. In this paper we provide 22 examples of how to compress solutions to combinatorial problems.

For the {\sc k-Sat}, for every variable size $n$, there are {\sc k-Sat}
instances which admit a single solution $S$. By permutating the variables and applying proper negations to this single solution $S\in\BT^n$, where $S[i]=1$ iff variable $x_i$ of the solution is \textit{true}, $S$ can be transformed so it cannot be compressed, with $n< \K(S)+O(1)$, where $\K$ is the prefix-free Kolmogorov complexity. However given more assumptions, and by using derandomization techinques, introduced in  \cite{EpsteinDerandom22}, one can improve upon the bounds for compressing solutions to {\sc k-Sat}. By Theorem \ref{thr:ksat} in Section \ref{sec:ksat},\\

\noindent\textbf{Theorem.} \textit{ Let $\phi$ be a {\sc k-Sat} instance of $n$ variables and $m$ clauses, with $k\geq 3$. If each clause intersects at most $(2^k/e)-1$ other clauses, then there exists a satisfying assignment $\psi$ of $\phi$ of complexity $\K(\psi)\lel \K(n)+2em/2^k +\I(\phi;\ch)$. }\\

The term $\I(\phi;\ch)$ is the amount of information that $\phi$ has with the halting sequence, defined in Section \ref{sec:tool}. Another area covered in this paper are examples of the compression size of players in interactive games. Take for example the {\sc Even-Odds} game in Section \ref{sec:EvenOdds}. At every round, the player and environment each play a 0 or a 1. If the sum is odd the player gets a point. If the sum is even the player loses a point. The player can't see the environment's actions but the environment can see the player's actions. The environment can have its actions be a function of the player's previous actions. By using derandomization, in Section \ref{sec:EvenOdds}, we prove the following result.\\

\noindent\textbf{Theorem.} \textit{Playing {\sc Even-Odds} for large enough $N$ rounds, for every environment $\q$ there is a deterministic agent $\p$ that can achieve a score of $\sqrt{N}$and $\K(\p)\lel \I(\q;\ch)$.}\\

The bounds in this paper are achieved using derandomization. This process entails showing that there exists a simple probability measure over the solution-candidate space such that the solutions have a  large enough probability. In the next step, Theorem \ref{thr:el} is used to show that there exists a simple solution based on this probability.

\tableofcontents
\section{Tools}
\label{sec:tool}
Let $\N$, $\R$, $\BT$, and $\FS$, consist of natural numbers, reals, bits, and finite sequences. $\R_{\geq 0}$ denotes nonnegative reals. $\|x\|$ is the length of a string. For mathematical statement $A$, $[A]=1$ if $A$ is true, otherwise $[A]=0$.  For $x\in\FS$ and $y\in\FS$, we use $x\sqsubseteq y$ if there is some string $z\in\FS$ where $xz=y$. We say $x\sqsubset y$ if $x\sqsubseteq y$ and $x\neq y$.

For positive real functions $f$ the terms  ${\lea}f$, ${\gea}f$, ${\eqa}f$ represent ${<}f{+}O(1)$, ${>}f{-}O(1)$, and ${=}f{\pm}O(1)$, respectively. The term ${\eqm}f$  denotes ${\lem}f$ and ${\gem}f$. For the nonnegative real function $f$, the terms ${\lel}f$, ${\gel} f$, and ${\eql}f$ represent the terms ${<}f{+}O(\log(f{+}1))$, ${>}f{-}O(\log(f{+}1))$, and ${=}f{\pm}O(\log(f{+}1))$, respectively. 

The function $\K(x|y)$ is the conditional prefix Kolmogorov complexity. By the chain rule $\K(x,y)\eqa \K(x)+\K(y|x,\K(x))$. The universal probability of a string $x\in\FS$, conditional on $y\in\FS\cup\IS$, is $\m(x|y)=\sum\{2^{-\|p\|}:U_y(p)=x\}$. Given any lower-computable semi-measure $P$ over $\FS$, $D\subseteq \FS$, $O(1)\m(D) > 2^{-\K(P)}P(D)$. The coding theorem states $-\log \m(x|y)\eqa\K(x|y)$.  The mutual information of a string $x$ with the halting sequence $\mathcal{H}\in\IS$ is $\I(x;\mathcal{H})=\K(x)-\K(x|\mathcal{H})$.

\begin{lmm}[Symmetric Lovasz Local Lemma]
\label{lmm:lll}
Let $E_1,\dots, E_n$ be a collection of events such that $\forall i:\Pr[E_i]\leq p$. Suppose further that each event is dependent on at most $d$ other events, and that $ep(d + 1) \leq 1$. Then, $\Pr\left[\bigcap_i\overline{E}_i\right]>\left(1-\frac{1}{d+1}\right)^n$.
\end{lmm}

\begin{lmm}[\cite{EpsteinDerandom22}]
	\label{lmm:consH}
		For partial computable $f:\N\rightarrow\N$, for all $a\in\N$, $\I(f(a);\mathcal{H})\lea\I(a;\mathcal{H})+\K(f)$. 
\end{lmm}
\begin{prf}
	\begin{align*}
	\I(a;\ch)&=\K(a)-\K(a|\ch)\gea \K(a,f(a))-\K(a,f(a)|\ch)-\K(f).
	\end{align*}
	The chain rule ($\K(x,y)\eqa \K(x)+\K(y|x,\K(x))$) applied twice results in
	\begin{align*}
	\I(a;\ch)+\K(f)&\gea \K(f(a))+\K(a|f(a),\K(f(a)))-(\K(f(a)|\ch)+\K(a|f(a),\K(f(a)|\ch),\ch)\\\
	&\eqa \I(f(a);\ch)+\K(a|f(a),\K(f(a)))-\K(a|f(a),\K(f(a)|\ch),\ch)\\
	&\eqa  \I(f(a);\ch)+\K(a|f(a),\K(f(a)))-\K(a|f(a),\K(f(a)),\K(f(a)|\ch),\ch)\\
	&\gea \I(f(a);\ch).
	\end{align*}
	\qed
\end{prf}

\begin{thr}[\cite{Levin16,Epstein19}]
\label{thr:el}$ $\\
For finite $D\subset\FS$, $-\log\max_{x\in D}\m(x)\lel -\log\sum_{x\in D}\m(x)+\I(D;\ch)$.
\end{thr}

A continuous semi-measure $Q$ is a function $Q:\FS\rightarrow \R_{\geq 0}$, such that $Q(\emptyset) =1$ and for all $x\in\FS$, $Q(x)\geq Q(x0)+Q(x1)$. For prefix free set $D$, $Q(D)=\sum_{x\in D}Q(x)$.  Let $\M$ be a largest, up to a multiplicative factor, lower semi-computable continuous semi-measure. That is, for all lower computable continuous semi-measures $Q$ there is a constant $c\in\N$ where for all $x\in\FS$, $c\M(x)>Q(x)$. Thus for any lower computable continuous semi-measure $W$ and prefix-free set $S\subset\FS$, $-\log\M(S)\lea \K(W) -\log W(S)$, where $\K(W)$ is the size of the smallest program that lower computes $W$. The monotone complexity of a finite prefix-free set $G$ of finite strings is $\Km(G)\edf \min\{\|p\|\,{:}\; U(p)\in x\sqsupseteq y\in G\}$. Note that this differs from the usual definition of $\Km$, in that our definition requires $U$ to halt. 

\begin{thr}[\cite{EpsteinDerandom22}]
\label{thr:mon}
	For finite prefix-free set $G\subset\FS$, we have $\Km(G)\lel -\log\M(G)+\I(G;\ch)$.
\end{thr}

Derandomization can also be applied to games (see \cite{Hutter05}), which consists of an interaction between an agent and an environment. In this paper we use two versions of the cybernetic agent model. For the first model, the agent $\p$ and environment $\q$ are defined as follows. The agent is a function $\p:(\N\times\N)^*\rightarrow\N$, where if $\p(w)=a$, $w\in (\N\times\N)^*$ is a list of the previous actions of the agent and the environment, and $a\in\N$ is the action to be performed. The environment is of the form $\q:(\N\times\N)^*\times\N\rightarrow \N\cup\{\mathbf{W}\}$, where if $\q(w,a)=b\in\N$, then $b$ is $\q$'s response to the agent's action $a$, given history $w$, and the game continues. If $\q$ responds $\mathbf{W}$ then the agents wins and the game halts. The agent can be randomized but the environment cannot be. The game can continue forever, given certain agents and environments. This is called a win/no-halt game.

\begin{thr}[\cite{EpsteinDerandom22}]
\label{thr:probwin}
If probabilistic agent $\p'$ wins against environment $\q$ with at least probability $p$, then there is a deterministic agent $\p$ of complexity $\lel \K(\p') -\log p + \I(( p, \p',\q);\ch)$ that wins against $\q$.
\end{thr}

The second game is modified such that the environment gives a nonnegative rational penalty term to the agent at each round. Furthermore the environment specifies an end to the game without specifying a winner or loser. This is called a penalty game.

\begin{cor}[\cite{EpsteinDerandom22}]
\label{cor:pengame}
If given probabilistic agent $\p$, environment $\q$ halts with probability 1, and 
 $\p$ has expected penalty less than $n\in\N$, then there is a deterministic agent of complexity $\lel \K(\p) +\I(( \p,n,\q);\ch)$ that receives penalty $<2n$ against $\q$.
\end{cor}

\section{Examples}
In this section 22 examples of derandomization are given. Some use the Lovasz Local Lemma, which is particularly suited for derandomization. There are 4 instances of games.
 \subsection{{\sc k-Sat}}
 \label{sec:ksat}
 
 \begin{figure}
 \fontsize{16}{15}
        \centering
       {{\sc $\left(x_1\vee x_3\vee \overline{x}_5\right)\wedge \left(\overline{x}_1\vee x_2 \vee \overline{x_3}\right)\wedge\left(\overline{x_2}\vee \overline{x}_4\vee\overline{x}_5\right)$}}
  \fontsize{12}{15}
         \caption{An example {\sc 3-Sat} instance. Each clause contains 3 literals consisting of variables $x_i$ or their negations $\overline{x}_i$. An example satisfying assignment is $x_1=\mathrm{True}, x_2=\mathrm{True}, x_3=\mathrm{False}, x_4=\mathrm{False}, x_5 = \mathrm{True}$.}
         \label{fig:ksat}\end{figure}
 For a set of $n$ Boolean variables $x_1,\dots,x_n$, a \textit{CNF} formula $\phi$ is a conjunction $C_1\cap\dots \cap C_m$ of clauses. Each clause $C_j$ is a disjunction of $k$ literals, where each literal is a variable $x_i$ or its negation $\overline{x_i}$. Clauses $C_j$ and $C_l$ are said to intersect if there is some $x_i$ such that both clauses contain either $x_i$ or $\overline{x_i}$. A satisfying assignment is a setting of each $x_i$ to true or false that makes $\phi$ evaluate to true. An example of {\sc k-Sat} can be seen in Figure \ref{fig:ksat}.
 \begin{thr}
 \label{thr:ksat}
 Let $\phi$ be a {\sc k-Sat} instance of $n$ variables and $m$ clauses, with $k\geq 3$. If each clause intersects at most $(2^k/e)-1$ other clauses, then there exists a satisfying assignment $\psi$ of $\phi$ of complexity $\K(\psi)\lel \K(n)+2em/2^k +\I(\phi;\ch)$. 
 \end{thr}
\begin{prf}
The sample space is the set of all $2^n$ assigments, and for each clause $C_J$, $E_j$ is the bad event ``$C_j$ is not satisfied''. Let $p=2^{-k}$ and $d=(2^k/e)-1$. Thus $\forall j$, $\Pr[E_j]\leq p$ as each clause has size $k$ and each $E_j$ is dependent on at most $d$ other events by the intersection property. Thus since $ep(d+1)$, by the Lovasz Local Lemma \ref{lmm:lll}, we have that, 
\begin{align}
\label{eq:ksat}
\Pr\left[\bigcap_j\overline{E_j}\right]>\left(1-\frac{1}{d+1}\right)^m=\left(1-\frac{e}{2^k}\right)^m.
\end{align}
Let $D\subset\BT^n$ be the set of all assignments that satisfy $\phi$. $\K(D|\phi)=O(1)$. Let $P$ be the uniform measure over sequences of size $n$. By Equation \ref{eq:ksat}, assuming $k\geq 3$,
$$-\log P(D)< -m\log(1-e/2^k)< 2em/2^k.$$
Thus by Theorem \ref{thr:el} and Lemma \ref{lmm:lll}, for $k\geq 3$, there exists an assignment $\psi\in D$ that satisfies $\phi$ with complexity
$$\K(\psi)\lel \K(P)-\log P(D)+\I(D;\ch)\lel \K(n)+2em/2^k+\I(\phi;\ch).$$\qed
\end{prf}

\subsection{{\sc Hypergraph-Coloring}}
In this section we show how to compress colorings of $k$-uniform hypergraph. A \textit{hypergraph} is a pair $J=(V,E)$ of vertices $V$ and edges $E\subseteq \mathcal{P}(V)$. Thus each edge can connect $\geq 2$ vertices. A hypergraph is $k$-\textit{uniform} of the size $|e|=k$ for all edges $e\in E$. A 2-uniform hypergraph is just a simple graph. A valid $C$-\textit{coloring} of a hypergraph $(V,E)$ is a mapping $f:V\rightarrow \{1,\dots,C\}$ where every edge $e\in E$ is not \textit{monochromatic} $|\{f(v):v\in e\}|>1$. The goal of {{\sc Hypergraph-Coloring}} with parameter $k$, is given a $k$ uniform hypergraph, produce a coloring using the smallest amount of colors. Theorem \ref{thr:hyper1} uses the union bound whereas Theorem \ref{thr:hyper2} uses the Lovasz Local Lemma.
\begin{thr}
\label{thr:hyper1}
Every $k$-uniform hypergraph $J=(V,E)$, $|E|=n$, $|V|=m$ has a $\ceil{\sqrt[\leftroot{-2}\uproot{2}k-1]{2m}}$ coloring $g$ where $\K(g)\lel \K(k,n,m)+ \I(J;\ch)$.
\end{thr} 
\begin{prf}
We randomly color every vertex $v\in V$ using $C=\ceil{\sqrt[\leftroot{-2}\uproot{2}k-1]{2m}}$
colors. Let $A_e$ be the bad event that edge $e$ is monochromatic. This event has probability:
$$
\Pr[A_e]=C\cdot(1/C)^{k}=(1/C)^{k-1}<1/2m,
$$
because there are $C$ possible colors and each vertex has a $1/C$ chance of getting a particular color. We can get a union-bound over all $m$ edges to find the bad probability.
\begin{align}
\label{eq:hyper1}
\Pr\left[\bigcup_{e\in E}A_e\right]<\sum_{e\in E}\Pr[A_e]<m\cdot(1/2m)=1/2.
\end{align}
We let $D\subset \BT^{n\ceil{\log C}}$ be the set of all encodings of  $C$ colorings (so no edge in monochromatic. $\K(D|J)=O(1)$. Let $P:\FS\rightarrow\R_{\geq 0}$ be a probability measure over $\FS$, uniformly distributed over all $x\in \BT^{n\ceil{\log C}}$ that encode a $C$ color assignment. $P(D)>.5$. By Theorem \ref{thr:el} and Lemma \ref{lmm:consH}, there is a graph coloring $g\in D$ where
$$
\K(g)\lel \K(P)-\log P(D) + \I(D;\ch)\lel \K(k,m,n) + \I(J;\ch).
$$\qed
\end{prf}
$ $\\
The second result on $k$-hypergraph coloring uses Lovasz Local Lemma.

\begin{thr}
\label{thr:hyper2}
Let $J=(V,E)$, $|V|=n$, $|E|=m$ be a hypergraph and $k=\min_{f\in E}|f|$, with $k\geq 3$. Assume for each edge $f$, there are at most $2^{k-1}/e$ edges $h\in E$ such that $h\cap f\neq\emptyset$. Then for $k\geq 4$, there is a 2-coloring $g$ of $G$ such that $\K(g)\lel \K(n) 4me/2^k + \I( J;\ch)$.
\end{thr}
 \begin{prf}
 The case is degenerate for $k=1$. Assume $k\geq 3$. We will use the Lovasz Local Lemma to get a lower bound on the probability that a random assignment of colors is a 2 coloring. We assume each vertex is colored black or white with equal probability. For each edge $f\in E$, we define $E_f$ to be the bad event ``$f$ is monochromatic''. A valid 2-coloring exists iff $\Pr\left[\bigcap_f\overline{E}_f\right]>0$.
 
  Let $p=1/2^{k-1}$ and $d=(2^{k-1}/e)-1$. For each $f$, $\Pr[E_f]\leq p$ by the fact that $f$ contains at least $k$ vertices. Furthermore since $f$ intersects at most $d$ edges besides itself, $E_f$ is dependent on at most $d$ of the other events. Therefore since $ep(d+1)=1$ we can apply the Lovasz Local Lemma \ref{lmm:lll},
  \begin{align}
  \label{eq:hyper2}
  &\Pr\left[\bigcap_{f}\overline{E}_f\right]>\left(1-\frac{1}{1+d}\right)^m=\left(1-\frac{e}{2^{k-1}}\right)^m.
  \end{align}
 
 Let $D=\BT^n$ be the set of all encoded 2 colorings of $J$. $\K(D|J)=O(1)$. Let $P(x)=[\|x\|=n]2^{-n}$ is the uniform distribution over sequences of length $n$. By Equation \ref{eq:hyper2}, assuming $k\geq 4$
 \begin{align*}
 -\log P(D) &<-m\log(1-2e/2^k)<4me/2^k.
 \end{align*}
 By Theorem \ref{thr:el} and Lemma \ref{lmm:lll}, there exist a 2-coloring $g$ of $J$ such that for $k\geq 4$,
 $$\K(g)\lel \K(P)-\log P(D) + \I(D;\ch) \lel \K(n)+4me/2^k+\I(J;\ch).$$
 \qed
 \end{prf}
 
\subsection{{\sc Vertex-Disjoint-Cycles}}
\begin{figure}     
  \centering
  \includegraphics[width=6cm]{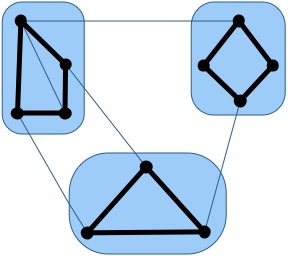}
  \caption{An example graph partitioned into 3 groups such that each group contains a cycle.}
  \label{fig:disjoint}
\end{figure}

\begin{prp}[Mutual Independence Principle]
\label{prp:mip}
Suppose that $Z_1,\dots Z_m$ is an underlying sequence of independent events and suppose that each event $A_i$ is completely determined by some subset $S_i\subset\{Z_1,\dots,Z_m\}$. If $S_i\cap S_j=\emptyset$ for $j=j_1,\dots,j_k$ then $A_i$ is mutually independent of $\{A_{j_1},\dots,A_{j_k}\}$.
\end{prp}
This section deals with partitioning graphs into subgraphs such that each subgraph contains an independent cycle. An example partition can be seen in Figure \ref{fig:disjoint}.
\begin{thr}
There is a partition $\ell$ of vertices of a $k$-regular graph $G=(V,E)$, with vertices $|V|=n$, into $c=\floor{\frac{k}{3\ln k}}$ components each containing a cycle that is vertex disjoint from the other cycles with complexity $\K(\ell)\lel\K(n,k)+2n/k^2 +\I(G;\ch)$.
\end{thr}
\begin{prf}
We partition the vertices of $G$ into $c=\floor{k/3\ln k}$ components by assigning each vertex to a component chosen independently and uniformly at random. With positive probability, we show that every component contains a cycle. It is sufficient to prove that every vertex has an edge leading to another vertex in the same component. This implies that starting at any vertex there exists a path of arbitrary length that does not leave the component of the vertex, so a sufficiently long path must include a cycle. A bad event $A_v=\{\textrm{vertex $v$ has no neighbor in the same component}\}$. Thus
\begin{align*}
\Pr[A_v] &=\prod_{(u,v)\in E}\Pr[\textrm{$u$ and $v$ are in different components}]\\
&=\left(1-\frac{1}{c}\right)^k<e^{-k/c}\leq e^{-3\ln k}=k^{-3}.
\end{align*}x
$A_v$ is determined by the component choices of itself and of its out neighbors $N^{\mathrm{out}}(v)$ and these choices are independent. Thus by the Mutual Independence Principle, (Proposition \ref{prp:mip}) the dependency set of $A_v$ consist of those $u$ that share a neighor with $v$, i.e., those $u$ for which $(\{v\}\cup N(v))\cap(\{u\}\cup N(u))\neq 0$. Thus the size of this dependency is at most $d=(k+1)^2$.

Take $d=(k+1)^2$ and $p=k^{-3}$, so $ep(d+1)=e(1+(k+1)^2)/k^3\leq 1$, holds for  $k\geq 5$. One can trivially find a partition of a $k$-regular  graph when $k<5$ because $c=1$. Thus, noting that $k\geq 5$,
\begin{align}
\label{eq:hyper3}
\Pr\left[\bigcap_{v\in G}\overline{A}_v\right]&>\left(1-\frac{1}{d+1}\right)^n=\left(1-\frac{1}{(k+1)^2+1}\right)^n>\left(1-\frac{1}{k^2}\right)^n.
\end{align}
$$-\log P(D)< -n\log(1-1/k^2) < 2n/k^2.$$

By Theorem \ref{thr:el} and Lemma \ref{lmm:consH}, for large enough $k$, there exist a partitioning $\ell\in D$ of vertices into $c=\floor{k/3\ln k}$ components each containing a cycle that is vertex disjoint from the other cycles with complexity 
$$\K(\ell)\lel \K(P) -\log P(D)+\I(D;\ch)\lel \K(n,k)+2n/k^2 + \I(G;\ch).$$\qed
\end{prf}
\subsection{{\sc Weakly-Frugal-Graph-Coloring}}
For an undirected graph $G=(V,E)$, a $k$-coloring assignment $f:V\rightarrow\{1,\dots,k\}$ is a $\beta-$\textit{weakly frugal} if for all neighbors of vertices $v\in V$ contain at most $\beta$ vertices with the same assignment. Note that a weakly frugal coloring assignment differs from a frugal coloring assignment, introduced in \cite{HindMoRe97}, by the fact that the former can have two adjacent vertices with the same color.

\begin{thr}
For graph $G=(V,E)$, with $|V|=n$, with max degree $\Delta>2e$ there is a $\beta$-weakly frugal coloring assignment $f$, with $\beta<\Delta$, using $Q\geq \Delta^{1+4/\beta}/2$ colors with complexity $\K(f)\lel \K(n,Q) + 2n/\beta+\I((G,\beta,Q);\ch)$.
\end{thr}
\begin{prf}
This proof is a modification of the proof in \cite{HindMoRe97}, except the restriction is relaxed to weakly-frugal coloring. Let us say each vertex is assigned one of $Q$ colors with uniform randomness. For vertices $\{u_1,\dots,u_{\beta+1}\}$ that are in the neighborhood of a vertex $v\in V$, let $B_{u_1,\dots,u_{\beta+1}}$ be the bad event that the vertices are the same color. $\Pr[B_{u_1,\dots,u_{\beta+1}}]=p=1/Q^{\beta}$. Each such bad event is dependent on at most $d=(\beta+1)\Delta{\Delta\choose \beta}$ other events. There are at most $m=n{\Delta\choose \beta+1}$ such events. The requirement that $ep(d+1)\leq 1$ of the Lovasz Local Lemma is fulfilled, because
\begin{align*}
&ep(d+1)\\
=&e\frac{1}{Q^\beta}\left(1+(\beta+1)\Delta{\Delta\choose\beta}\right)\\
\leq&e\frac{1}{Q^\beta}\left(1+(\beta+1)(\Delta^{\beta+2}/\beta!)\right)\\
\leq&e\frac{1}{Q^\beta}(1+(\Delta^{\beta+3}/\beta!))\\
\leq&\frac{1}{Q^\beta}(\Delta^{\beta+4}/\beta!))\\
\leq&\frac{1}{Q^\beta}\Delta^{\beta+4}2^{-\beta}\\
\leq& 1.
\end{align*}

By Lovasz Local Lemma \ref{lmm:lll}, 
\begin{align*}
&-\log\Pr\left(\bigcap_{u_1,\dots,u_{\beta+1}}\overline{B}_{u_1,\dots,u_{\beta+1}}\right)\\
<&-m\log\left(1-\frac{1}{d+1}\right)\\
<&2m\left(1-\frac{1}{d+1}\right)\\
<&2n{\Delta\choose \beta+1}/\left(1+(\beta+1)\Delta{\Delta\choose \beta}\right)\\
<&2n{\Delta\choose \beta+1}/\left((\beta+1)\Delta{\Delta\choose \beta}\right)\\
<&2n/(\beta+1).\\
<&2n/\beta.\\
\end{align*}
Let $D\subset\BT^{n\ceil{\log Q}}$ be encodings of all $\beta$-weakly frugal coloring of $G$ using $Q$ colors. $\K(D|G,Q,\beta)=O(1)$. Let $P$ be uniform distribution over all $Q$-color assignments to $n$ vertices. $-\log P(D) \lea 2n/\beta$. By Theorem \ref{thr:el} and Lemma \ref{lmm:lll}, we have a $\beta$-weakly frugal color assignment $f\in D$ of $G$ using $Q$ colors such that
$$\K(f)\lel \K(P)-\log P(D) +\I(D;\ch)\lel \K(n,Q) +2n/\beta+\I((G,Q,\beta);\ch).$$
\qed
\end{prf}
\subsection{{\sc Graph-Coloring}}

\begin{figure}     
  \centering
  \includegraphics[width=6cm]{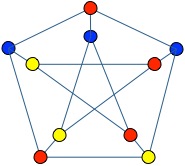}
  \caption{An example graph coloring. Nodes that share an edge are assigned different colors.}
  \label{fig:graphcoloring}
\end{figure}

For graph $G=(V,E)$, with undirected edges, a $k$-coloring is a function $f:V\rightarrow \{1,\dots,k\}$ such that if $(v,u)\in E$, then $f(v)\neq f(u)$. An example graph coloring can be seen in Figure \ref{fig:graphcoloring}

\begin{thr}
For graph $G=(V,E)$, $|V|=n$ with max degree $d$, there is a $k$ coloring $f$ with $2d\leq k$, and $\K(f) \lel \K(n,k)+2nd/k+ \I((G,k);\ch)$. 
\end{thr}
\begin{prf}
Let us say we randomly assign a color to each vertex. The probability that the color of the $i$th vertex does not conflict with the previous coloring is at least $(k-d)/k$. Thus the probability of a proper coloring is $\geq ((k-d)/k)^n$. Let $D\subseteq \BT^{n\ceil{\log k}}$ be all encoded proper $k$ colorings of $G$. $\K(D|G,k)=O(1)$. Let $P:\FS\rightarrow \R_{\geq 0}$ be a probability measure that is the uniform distribution over all possible color assignments. Thus, assuming $d/k\leq .5$,
$$
-\log P(D)\leq -n\log (1-d/k) \leq 2nd/k.
$$
Thus by Theorem \ref{thr:el} and Lemma \ref{lmm:consH}, there is a coloring $f\in D$ with 
\begin{align*}
\K(f)&\lel -\log \m(D) +\I(D;\ch)\\
&\lel \K(P)-\log P(D)+\I(D;\ch)\\
&\lel \K(n,k)+2nd/k + \I((G,k);\ch).
\end{align*}\qed
\end{prf}

\subsection{{\sc Max-Cut}}
Imagine a graph $G=(E,V)$, $|V|=n$, consisting of vertices $V$ and undirected edges $E$, and a weight $\omega_e$ for each edge $e\in \E$. Let $\omega=\sum_{e\in E}\omega_e$ be the combined weight of all edges. The goal is to find a partition $(A,B)$ of the vertices into two groups that maximizes the total weight of the edges between them.
\begin{thr}
There is a cut $f$ of $G$ that is $1/3$th optimal and $\K(f)\lel \K(n)+\I(G;\ch)$.
\end{thr}
\begin{prf}
 Imagine the algorithm that on receipt of a vertex, randomly places it into $A$ or $B$ with equal probability. Then the expected weight of the cut is
$$
\E\left[\sum_{e\in E(A,B)}\omega_e\right]=\sum_{e\in E}\omega_e\Pr(e\in E(A,B)) = \frac{1}{2}\omega.
$$
This means the expected weight of the cut is at least half the weight of the maximum cut. Some simple math results in the fact that $\Pr\left[\sum_{e\in E(A,B)}\omega_e\right]>\omega/3 \geq 1/4$. We can encode a cut into a binary string of $x$ length $n$, where $x[i]=1$, if the $i$th vertex is in $A$. Let $P$ be the uniform distribution over strings of size $n$. Let $D\subset\BT^n$ consist of all encoded cuts that are at least $1/3$ optimal. $\K(D|G)=O(1)$ and $P(D)\geq .25.$
By Theorem \ref{thr:el} and Lemma \ref{lmm:consH},
\begin{align*}
\min_{f\in D}\K(f) &\lel\K(P) -\log P(D)+\I(D;\ch)\lel \K(n)+\I(G;\ch).
\end{align*}
\qed
\end{prf}

\subsection{{\sc Max-3Sat}}
 This problem consists of a boolean formula $f$ in conjunctive normal form, comprised of $m$ clauses, each consisting of a disjunction of 3 literals. Each literal is either a variable or the negation of a variable. We assume that no literal (including its negation) appears more than once in the same clause. There are $n$ variables. The goal is to find an assignment of variables that satisfies as many clauses as possible.
\begin{thr}
There is an assignment $x$ that is $6/7$th optimal and has complexity $\K(x)\lel\K(m)+\I(f;\ch)$.
\end{thr} 
\begin{prf}
  The randomized approximation algorithm is as follows. The variables are assigned true or false with equal probability. Let $Y_i$ be the random variable that clause $i$ is satisfied. Thus the probability that clause $Y_i$ is satisfied is $7/8$. So the total expected number of satisfied clauses is $7m/8$, which is $7/8$ of optimal. Some simple math shows the probability that number of satified clauses is $>6m/7$ is at least $1/8$.

Let $x\in\BT^n$ encode an assignment of $n$ variables, where $x[i]=1$ if variable $i$ is true. Let $D\subset\BT^n$ encode all assignments that are $6/7$th optimal. Let $P$ be the uniform distribution over strings of length $n$. $\K(D|f)=O(1)$ and $P(D)\geq 1/8$. By Theorem \ref{thr:el} and Lemma \ref{lmm:consH},
$$
\min_{x\in D}\K(x)\lel \K(P)-\log P(D) + \I(D;\ch)\lel \K(n)+\I(f;\ch).
$$
\qed
\end{prf}
\subsection{{\sc Balancing-Vectors}}
For a vector $v=(v_1,\dots,v_n)\in\R^n$, $\|v\|_\infty=\max_i|V_i|$. Binary matrix $M$ is a matrix whose values are either 0 or 1. The goal of {\sc Binary Matrix}, is given $M$, to find a vector $b\in\{-1,+1\}^n$ that minimizes $\|Mb\|_\infty$.
\begin{thr}
Given $n\times n$ binary matrix $M$, there is a vector $b=\{-1,+1\}^n$ such that $\|Mb\|_\infty\leq 4\sqrt{n\ln n}$ and $\K(b)\lel \K(n)+\I( M;\ch)$.
\end{thr}
\begin{prf}
Let $v=(v_1,\dots,v_n)$ be a row of $M$. Choose a random $b=(b_1,\dots,b_n)\in\{-1,+1\}^n$. Let $i_1,\dots,i_m$ be the indices such that $v_{i_j}=1$. Thus
$$
Y=\langle v,b\rangle=\sum_{i=1}^nv_ib_i=\sum_{j=1}^mv_{i_j}b_{i_j}=\sum_{j=1}^mb_{i_j}.
$$
$$
\E[Y]=\E[\langle v,b\rangle]=\E\left[\sum_iv_ib_i\right]=\sum_i\E[v_ib_i]=\sum v_i\E[b_i]=0.
$$
By the Chernoff inequality and the symmetry $Y$, for $\tau=4\sqrt{n\ln n}$,
$$\Pr[|Y|\geq\tau]=2\Pr[v\cdot b\geq \tau]=2\Pr\left[\sum_{j=1}^mb_{i_j}\geq \tau\right]\leq 2\exp\left(-\frac{\tau^2}{2m}\right)=2\exp\left(-8\frac{n\ln n}{m}\right)\leq 2n^{-8}.
$$
Thus, the probability that any entry in $Mb$ exceeds $4\sqrt{n\ln n}$ is smaller than $2n^{-7}$. Thus, with probability $1-2n^{-7}$, all the entries of $Mb$ have value smaller than $4\sqrt{n\ln n}$.

Let $P:\FS\rightarrow \R_{\geq 0}$, be the uniform measure over string of length $n$, with $P(x)=[\|x\|=n]2^{-n}$. Let $D$ consist of all strings that encode vectors $b_x\in\{-1,+1\}^n$ in the natural way such that $\|Mb_x\|_{\infty}\leq 4\sqrt{n\ln n}$. $\K(D|M)=O(1)$. Thus by the above reasoning $P(D)\geq 1- 2n^{-7} >0.5$. By Theorem \ref{thr:el} and Lemma \ref{lmm:consH}, there exists an $x\in D$, such that
$$\K(x)\lel \K(P)-\log P(D)+\I(D;\ch)\lel \K(n)+\I(M;\ch).$$
Thus there exists a $b_x\in\{-1,+1\}^n$ that satisfies the theorem statement.\qed
\end{prf}

We provide another derandomization example using balancing vectors.
\begin{thr}
Let $v=v_1,\dots,v_n\in\R^n$, all $|v_i|=1$, Then there exist $\epsilon=\epsilon_1,\dots,\epsilon_n=\pm 1$ such that $|\epsilon_1v_1+\dots+\epsilon_nv_n|\leq \sqrt{2n}$ and $\K(\{\epsilon\}) \lel \K(n)+\I(v;\ch)$.
\end{thr}
\begin{prf}
Let $\epsilon_1,\dots,\epsilon_n$ be selected uniformly and independently from $\{-1,+1\}$. Set
$$
X=|\epsilon v_1+\dots+\epsilon_nv_n|^2.
$$
Then 
$$
X=\sum_{i=1}\sum_{j=1}^n\epsilon_i\epsilon_jv_i\cdot v_j.
$$
So
$$
\E[X]=\sum_{i=1}^n\sum_{j=1}^nv_i\cdot v_j\E[\epsilon_i\epsilon_j]
$$
When $i\neq j$, $\E[\epsilon_i\epsilon_j]=\E[\epsilon_i]\E[\epsilon_j]=0$. When $i=j$, $\E[\epsilon^2_i]=1$, so
$$
\E[X]=\sum_{i=1}^nv_i\cdot v_i = n$$
So $\Pr[X\leq 2n] \geq 0.5$. Let $D\subseteq \BT^n$ consist of sequences of length $n$, each encoding an assignment of $\epsilon_1$ to $\epsilon_n$ in the natural way, such that the assignment of $\epsilon$ results in an $X_\epsilon
\leq 2n$. $\K(D|v)=O(1)$. Let $P$ be the uniform measure over sequences of length $n$. By the above reasoning $P(D)\geq 0.5$. By Theorem \ref{thr:el} and Lemma \ref{lmm:consH}, there is an assignment $\epsilon\in D$, such that
$\K(\epsilon)\lel \K(P)-\log P(D)+\I(D;\ch)\lel \K(n)+\I(v;\ch)$.
This assignment has $X_\epsilon\leq 2n$. Thus $|\epsilon_1v_1+\dots\epsilon_nv_n|\leq \sqrt{2n}$, satisfying the theorem.\qed
\end{prf}

\subsection{{\sc Parallel-Routing}}
The {\sc Parallel-Routing} problem consists of $(G,d)$, a directed graph $G=(N,V)$ and a set of destinations $d:N\rightarrow N$. Each node represents a processor $i$ in a network containing a packet $v_i$ destined for another processor $d(i)$ in the network. The packet moves along a route represented by a path in $G$. During its transmission, a packet may have to wait at an intermediate node because the node is busy transmitting another packet. Each node contains a separate queue for each of its links and follows a FIFO queuing disciple to route packets, with ties handled arbitrarily. The goal of {\sc Parallel-Routing} is to provide $N$ routes from $i\in N$ to $d(i)$ that minimize lag time.

\begin{figure}     
  \centering
  \includegraphics[width=6cm]{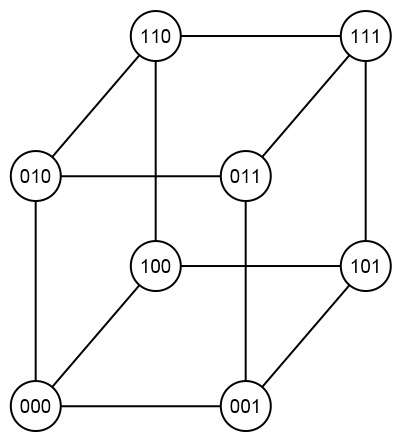}
  \caption{A Boolean Hypercube network, for $n=3$.}
  \label{fig:Routing}
\end{figure}

We restrict graphs to \textit{Boolean Hypercube} networks, which is popular for parallel processing. The cube network contains $N=2^n$ processing elements/nodes and is connected in the following manner. if $(i_0,\dots,i_{n-1})$ and $(j_0,\dots,j_{n-1})$ are binary representation of node $i$ and node $j$, then there exist directed edges $(i,j)$ and $(j,i)$ between the nodes if and only if the binary representation differ in exactly one position. An example Boolean Hypercube can be found in Figure \ref{fig:Routing}.

One set of solutions, called \textit{oblivious algrithms} satisfies the following property: a route followed by $v_i$ depends on $d(i)$ alone, and not on $d(j)$ for any $j\neq i$. We focus our attention on a 2 phase oblivious routing algorithm, {\sc Two-Phase}. Under this scheme, packet $v_i$ executes the following two phases independently of all the other packets.
\begin{enumerate}
\item Pick a intermediate destination $\sigma(i)$.  Packet $v_i$ travels to node $\sigma(i)$.
\item Packet $v_i$ travels from $\sigma(i)$ to destination $d(i)$.
\end{enumerate}
The method that the routes use for each phase is the \textit{bit-fixing} routing strategy. Its description is as follows.  To go from $i$ to $\sigma(i)$: one scans the bits of $\sigma(i)$ from left to right, and compares them with $i$. One sends $v_i$ out of the current node along the edge corresponding to the left-most bit in which the current position and $\sigma(i)$ differ. Thus going from $(1011)$ to $(0000)$, the packet would pass through $(0011)$ and then $(0001)$.

\begin{thr}
Given a {\sc Parallel-Routing} instance $(G,d)$, there is a set of intermediate destinations $\sigma:\N\rightarrow \BT^n$ for each $i$ such that every packet $i$ using $\sigma(i)$ and the {\sc Two-Phase} algorithm reaches its destination in at most $14n$ steps and $\K(\sigma)\lel \I( G,d;\ch)$.
\end{thr}
\begin{prf}

By Theorem 47 in \cite{MotwaniRa95}, if the intermediate destinations are chosen randomly, with probability least $1-(1/N)$, every packet reaches its destination in $14n$ or fewer steps. Let $D\subset\BT^{nN}$ be the set of all  intermediate destinations $\sigma\in D$ such that the lag time of instance $(G,d)$ using $\sigma$ is $\leq 14n$.  Let $\mu:\FS\rightarrow \R_{\geq 0}$ be the uniform continuous semi-measure, with $\mu(\emptyset)=1$, $\mu(x)= 2^{-\|x\|}$.
Thus $\mu(D)\geq 0.5$. $\K(D|(G,d))=O(1)$. Theorem \ref{thr:mon} and Lemma \ref{lmm:consH} results in 
$$\Km(D)\lel -\log \M(D)+\I( D;\ch)\lel -\log \mu(D)+\I((G,d);\ch)\lel\I((G,d);\ch).$$
Thus using $y\sqsupseteq x \in D$ that realizes $\Km(D)$, one can construct a function $\sigma:\N\rightarrow \BT^n$ which produces the desired intermediate destinations, and $\K(\sigma)\lea \K(y)\lel\I((G,d);\ch)$.\qed
\end{prf}

\subsection{{\sc Independent-Set}}
\begin{figure}     
  \centering
  \includegraphics[width=6cm]{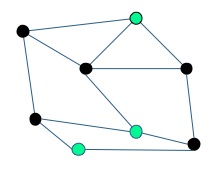}
  \caption{A graphical depiction of an independent set, represented by the green vertices. They do not share any edges.}
  \label{fig:ind}
\end{figure}

An independent set in a graph $G$ is a set of vertices with no edges between then, as shown in Figure \ref{fig:ind}. The {\sc Independent-Set} problem consists of an undirected graph $G$ and the goal is to find the largest independent set of that $G$.

\begin{thr}
For a graph $G$ on $n$ vertices with $m$ edges, there exists an independent set $S$ of size $0.75\sqrt{n}-2m/n$ and complexity $\lel \K(n,m)+4(\log n)(m/n)+\I(G;\ch)$.
\end{thr}
\begin{prf}
We use a modification of the algorithm in the proof of Theorem 6.5 in \cite{MitzenmacherUp05}. The randomized algorithm $A$ is as follows.
\begin{enumerate}
\item Delete each vertex (along with its incident edges) independently with probability $1-p$.
\item For each remaining edge, remove it and one of its adjacent vertices.
\end{enumerate}
For $X$, the number of vertices that survive the first round $\E[X]=np$. Let $Y$ be the number of edges that survive the first step, $\E[Y]=mp^2$. The second steps removes at most $Y$ vertices. The output is an independent set of size at least $\E[X-Y]=np-mp^2$. Let $p=1/\sqrt{n}$. Thus $\E[X] = \sqrt{n}$, $\E[Y]= m/n$, and $\E[X-Y]=\sqrt{n}-m/n$. By the Markov inequality, $\Pr[Y<2m/n]> 1/2$. By the Hoeffding's inequality, $$\Pr[X\leq  0.75\sqrt{n}]\leq e^{-2*(0.75)^2(np)^2/n}\leq e^{-2(.75^2)(n*n^{-.5})^2/n}\leq e^{-2*0.5}=e^{-1}.$$

For a sequence $x\in\FS$, $x_{[1]}= |\{i:x[i]=1\}|$ and $x_{[0]}=\|x\|-x_1$. Let $P:\FS\rightarrow \R_{\geq 0}$ be a computable probability, where for a string $x\in\BT^n$, $P(x)=(1/\sqrt{n})^{x_{[1]}}(1-1/\sqrt{n})^{x_{[0]}}$. Thus each $x$ represents a selection of vertices selected according to the randomized algorithm $A$. Let $D\subseteq\BT^n$ be the set consists of all sequences $x$ such that the $X$ variable resultant from $x$ is $|X_x| > 0.75\sqrt{n}$ and the $Y$ variable resultant from algorithm $A$ is $|Y_x|\leq 2m/n$. Thus $P(D) \geq (1-e^{-1})+1/2 -1 > 1/10$. Furthermore $D$ can be constructed from $G$, with $\K(D|G)=O(1)$. By Theorem \ref{thr:el} and Lemma \ref{lmm:consH}, there exists an $x\in D$, with
\begin{align*}
\K(x) &\lel \K(P) -\log P(D)+\I(D;\ch)\\
&\lel \K(n) + \I(G;\ch).
\end{align*}
In order for $x$ to represent an independent set, the second step of algorithm $A$ needs to be applied. In this case there are $<2m/n$ vertices that needs to be removed. Thus a modification $x'$ that has these vertices deleted represents an independent set. 
\begin{align*}
\K(x')\lel& \K(x,n,m) +(2\log n)(2m/n)\\
\lel& \K(n,m) + (4\log n)(m/n) + \I(D;\ch)\\
\lel& \K(n,m) + (4\log n)(m/n) + \I(G;\ch).
\end{align*}
This independent set has $X_x> 0.75\sqrt{n}$ and $Y_x<2m/n$, it size is $\geq 0.75\sqrt{n}-2m/n$.\qed
\end{prf}

\subsection{{\sc Dominating-Set}}
\begin{figure}     
  \centering
  \includegraphics[width=6cm]{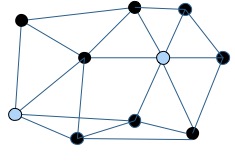}
  \caption{A graphical depiction of a dominating set. The two highlighted vertices are adjacent to all other vertices in the graph.}
  \label{fig:dom}
\end{figure}

A \textit{dominating-set} of an undirected graph $G=(E,V)$ on $n$ vertices is a set $U\subseteq V$ such that every vertex $v\in V-U$ has at least one neighbor in $U$. An example of a dominating set can be seen in Figure \ref{fig:dom}.

\begin{thr}
Every graph $G=(V,E)$, $|V|=n$ with min degree $\delta>1$ has a dominating set $U$ of size $\leq 3n\frac{1+\ln(\delta+1)}{\delta+1}$ and complexity $\K(U)\lel\K(n,\delta)+ 6(n\log n)/(\delta+1)+\I(G;\ch)$.
\end{thr}
\begin{prf}
Let $p\in[0,1]$. Let the vertices of $V$ be picked randomly and independently, each with probability $p$. Let $X$ be the random set of all vertices picked.  $\E[|X|]=np$. Let $Y=Y_X$ be the random set of all vertices $V-X$ that do not have a neighbor in $X$. $\Pr(v\in Y_X)\leq (1-p)^{\delta+1}$. Thus $\E[|Y_X| \leq n(1-p)^{\delta+1}\leq ne^{-p(\delta+1)}$.
 We set $p= \ln(\delta+1)/(\delta+1)$.  $\Pr[X\leq 3n\ln(\delta+1)/(\delta+1)]\geq 2/3$. $\Pr[Y_X \leq 3n/(\delta+1)]\geq 2/3$. Thus the probability of the previous two events is $\geq 1/3$.

Let $D\subseteq\BT^n$ be the set consisting of all sequences $x\in\BT^n$ where $x[i]=1$ indicates vertex $i$ was selected, such that the $X$ variable resultant from $x$ is $|X_x|\leq 3n\ln(\delta+1)/(\delta+1)]$ and the $Y_{x}$ resultant variable is $|Y_x|\leq 3n/(\delta+1)$. Furthermore $D$ can be constructed from $G$, with $\K(D|G)=O(1)$. Let $P:\FS\rightarrow\R_{\geq 0}$ be a probability measure over $x\in\BT^n$, where $P(x)=\prod_{i=1}^n\left( px[i]+(1-p)(1-x[i])\right)$. By definition of $D$, $P(D)\geq 1/3$. Furthermore by Theorem \ref{thr:el} and Lemma \ref{lmm:consH}, there is a subset of vertices $x\in D$, $x\subseteq V$, with
\begin{align*}
\K(x) &\lel \K(P)-\log P(D) +\I(D;\ch)\lel \K(n,\delta) + \I(G;\ch).
\end{align*}
The sequence $x$ represent the first step, however the set $Y_x$ needs to be added to make $x$ a dominating steps. Thus $3n/(\delta+1)$ vertices needs to be added, each can be encoded by $(2\log n)$ bits. Thus a dominating set $x'$ of $G$ exists of size  $\leq 3n\frac{1+\ln(\delta+1)}{\delta+1}$ such that 
$$\K(x')\lel \K(n,\delta) + 6(n\log n)/(\delta+1)+\I(G;\ch).$$\qed
\end{prf}

\subsection{{\sc Set-Membership}}
 For a set $G\subseteq\BT^\ell$, a function $f:\FS\rightarrow \BT$ is a partial checker for $G$, if $f(x)=1$ if $x\in G$. We use $\mathcal{U}$ to denote the uniform distribution over $\BT^\ell$. $\mathrm{Error}(G,f) = \Pr_{x\sim\mathcal{U}}[f(x)=1,x\not\in G]$.
 The goal of {\sc Set-Membership}, is given a set $G\subseteq\BT^\ell$, what is the simplest partial checker $f$ for $G$ that reduces $\mathrm{Error}(G,f)$.
  
\begin{thr}
For large enough $n$, given $G\subseteq \BT^\ell$, $|G|=m$, there is a partial checker $f$ such that 
$\mathrm{Error}(f,G)\leq 0.878^{n/m}$ and $\K(f) \lel \K(n,k,\ell) + n + \I( (G,n,k);\ch)$.
\end{thr}
\begin{prf}
We derandomize the Bloom filter algorithm \cite{Bloom70}. Let there be $k$ random functions $h_i:\BT^\ell\rightarrow \{1,\dots,n\}$, where each $h_i$ maps each input $x\in\BT^\ell$ to its range with uniform probability. We start with a string $v=0^n$. For each member $x\in G$, and $i\in \{1,\dots,k\}$, $v[h_i(x)]$ is set to 1. Thus the functions $h_i$ serve as a way to test membership of $G$. An example of the Bloom filter can be seen in Figure 
\ref{fig:bloom}. \begin{figure}     
  \centering
  \includegraphics[width=12cm]{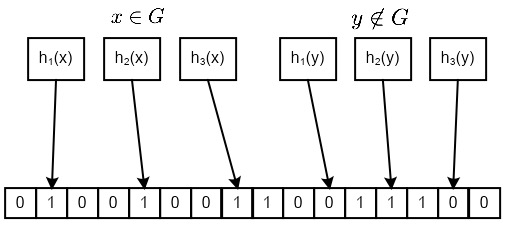}
  \caption{A graphical depiction of Bloom filter with $k=3$ hash functions. The first element $x$ is in $G$ and is thus mapped to ones in the Bloom filter. Thus the Bloom filter would indicate that $x\in G$. The second element $y$ is not in $G$ and has some of the hash functions map to 0. Thus the Bloom filter would indicate that $y\not\in G$.}
  \label{fig:bloom}
\end{figure} If $x\in G$, then all the indicator functions $h_i$ would be one. The probability that a specific bit is 0 is 
 $$p'=\left(1-\frac{1}{n}\right)^{km}.$$
 Let $X$ be the number of bins that are $0$. Due to \cite{MitzenmacherUp05},
 $$
 \Pr(|X-np'|\geq \epsilon n)\leq 2e\sqrt{n}e^{-n\epsilon^2/3p'}.
 $$
For $\epsilon=p'/10$, we get
\begin{align}
\label{eq:bloom1}
\Pr(X/n\geq p'9/10)&\leq 2e\sqrt{n}e^{-np'/300}.
\end{align}
Thus for proper choice of $k$ determined later, for large enough $n$, the right hand side of the above inequality is less than $0.5$. Thus with probability $>.5$, the expected false positive rate, $r$, that is $x\in\BT^\ell$, $x\not\in G$, $h_i(x)=1$, for all $i\in\{1,\dots,k\}$ is less than 
\begin{align*}
r\leq&(1-.9p')^k\\
=&\left(1-.9\left(1-\frac{1}{n}\right)^{km}\right)^k\\
\leq&\left(1-.9e^{-km/n}\right)^k.\\
\end{align*}

Setting $k=\ceil{n/m}$, with probability $\geq 1/2$, $r\leq (1-.5e^{-2})^{m/n}\leq 0.878^{m/n}$. Furthermore, for large enough $n$, $p'> .5e^{-\ceil{n/m}(m/n)}\geq .5e^{-2}$, which can be plugged back into Equation \ref{eq:bloom1}.

Let $F'\subset\FS$ consist of all encodings of $k$ hash functions $h_i:\BT^\ell\rightarrow \{1,\dots,n\}$. Let $F\subseteq F'$ consist of all hash functions such that the false positive rate $r$ is $\leq 0.878^{m/n}$. Let $P$ be the uniform distribution over $F'$. By the above reasoning, for large enough $n$, $P(F)>1/2$. $\K(F|G,k,n)=O(1)$. By Theorem \ref{thr:el} and Lemma \ref{lmm:consH}, there is an $h\in F$ such that
$$\K(h)\lel \K(P) - \log P(F) + \I( F;\ch)\lel \K(n,k,\ell)+\I( (G,n,k);\ch).$$ 

Thus $h$ represents a set of $k$ deterministic hash functions. Let $x$ be the Bloom filter using $h$ on $G$. Using $x$ and $h$, one can define a partial checker $f$ that is a Bloom filter such that $\mathrm{Error}(f,G)\leq 0.878^{n/m}$.  Furthermore,
$$\K(f)\lel \K(x,h)\lel \K(n,k,\ell)+n+\I( (G,n,k);\ch).$$
\qed
\end{prf}

\subsection{{\sc Latin-Transversal}}
\begin{figure}     
  \centering
  \includegraphics[width=4cm]{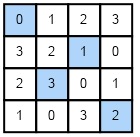}
  \caption{A graphical depiction of a Latin Transversal. Each number appears exactly 4 times in the matrix. The transversal is a permutation of the matrix such that all its entries have different values.}
  \label{fig:latin}
\end{figure}

Let $A=(a_i)$ be an $n\times n$ matrix with integer entries. A permutation $\pi$ is called a \textit{Latin Transversal} if the entries $a_{i\pi (i)}(1\leq i\leq n)$ are all distinct. An example Latin Transversal, where each integer occurs exactly 4 times, can be seen in Figure \ref{fig:latin}
\begin{lmm}[Lopsided Lovasz Local Lemma\cite{ErdosSp91}]
\label{lmm:alll}
Let $E_1,\dots, E_n$ be a collection of events with dependency graph $G=(V,E)$. Suppose $ \Pr\left(E_i\big|\bigcap_{j\in S}\overline{E_j}\right)\leq \Pr(E_i)$, for all $i,S\subset V$ with no $j\in S$ adjacent to $i$. Suppose all events have probability at most $p$, $G$ has degree at most $d$, and $4dp\leq 1$. Then $\Pr(\bigcap_i\overline{E_i})\geq (1-2p)^n$.
\end{lmm}
\begin{thr}
Suppose $k\leq (n-1)/16$ and suppose integers appears in exactly $k$ entries of $n\times n$ matrix $A$. Then for $n\geq 3$, $A$ has a Latin Traversal $\tau$ of complexity $\K(\tau)\lel \K(n) +4(k-1)+\I(A;\ch)$.
\end{thr}
\begin{prf}
Let $\pi$ be a random permutation $\{1,2,\dots,n\}$, chosen according to a uniform distribution $P$ among all possible $n!$ permutations. Define $T$ by the set of all ordered fourtuples $(i,j,i',j')$ with $i<i'$, $j\neq j'$, and $a_{ij}=a_{i'j'}$. For each $(i,j,i',j')\in T$, let $A_{iji'j'}$ denote the bad event that $\pi(i)=j$ and $\pi(i')=(j')$. Thus $A_{iji'j'}$ is the bad event that the random permutation has a conflict at $(i,j)$ and $(i',j')$. 

Clearly $P(A_{iji'j'})=1/n(n-1)$. The existence of a Latin Transversal is equivalent to the statement that with positive probability, none of these events hold. We define a symmetric digraph $G$ on the vertex set $T$ by making $(i,j,i',j')$ adjacent to $(p,q,p',q')$ if $\{i,i'\}\cap \{p,p'\}\neq\emptyset$ or $\{j,j'\}\cap\{q,q'\}\neq\emptyset$. Thus these two fourtuples are not adjacent iff the four cells $(i,j)$, $(i',j')$, $(p,q)$ and $(p',q')$ occupy four distinct rows and columns of $A$. 

The maximum degree of $G$ is less than $4nk\leq d$ because for a given $(i,j,i',j')\in T$ there are at most $4n$ choices of $(s,t)$ with either $s\in\{i,i'\}$ or $t\in\{j,j'\}$ and for each of these choices of $(s,t)$ there are less than $k$ choices for $(s',t')\neq (s,t)$ with $a_{st}=a_{s't'}$. Each fourtuple $(s,t,s',t')$ can be uniquely represented as $(p,q,p',q')$ with $p<p'$. Since $4dp\leq 16nk/(n(n-1)\leq 1$, by the Lopsided Lovasz Local Lemma, \ref{lmm:alll}, the desired bounds can be achieved if we can show that 
$$\Pr\left(A_{iji'j'}\big|\bigcap_S{A}_{pqp'q'}\right)\leq 1/n(n-1),$$
for any $(i,j,i',j')\in T$ and any subset $S$ of $T$ which are not-adjacent in $G$ to $(i,j,i',j')$. By symmetry we can assume $i=j=1$, $i'=j'=2$. A permutation $\pi$ is \textit{good} if it satisfies  $\bigcap_S\overline{A}_{pqp'q'}$ and let $S_{ij}$ denote the set of all good permutations $\pi$ satisfying $\pi(1)=i$ and $\pi(2)=j$. $|S_{12}|\leq |S_{ij}|$ for all $i\leq j$. 

Indeed suppose first that $i,j>2$. For each good $\pi\in S_{12}$ define a permutation $\pi^*$ as follows. Suppose $\pi(x)=i$ and $\pi(y)=j$. Then define $\pi^*(1)=i$, $\pi^*(2)=j$, $\pi^*(x)=1$, $\pi^*(y)=2$ and $\pi^*(t)=\pi(t)$ for all $t\neq 1,2,x,y$. One can easily check that $\pi^*$ is good, since the cells $(1,i),(2,j),(x,1),(y,2)$ are not part of any $(p,q,p',q')\in S$. Thus $\pi^*\in S_{ij}$ and since the mapping $\pi\rightarrow \pi^*$  is injective $|S_{12}|\leq |S_{ij}|$. One can define an injection mappings showing that $|S_{12}|\leq |S_{ij}|$ even when $\{i,j\}\cap\{1,2\}\neq\emptyset$. If follows that $\Pr\left(A_{1122}\cap \bigcap_S\overline{A}_{pqp'q'}\right)\leq \Pr\left(A_{1i2j}\cap\bigcap_S\overline{A}_{pqp'q'}\right)$ and hence $\Pr\left(A_{1122}\big|\bigcap_{S}\overline{A}_{pqp'q'}\right)\leq 1/n(n-1)$. 

The number of bad events $A_{iji'j'}$ is $\left(\frac{n^2}{k}\right){k\choose 2}$, as there are $n^2/k$ distinct numbers, and each number appears $k$ times. Thus by the Lopsided Lovasz Local Lemma \ref{lmm:alll}, for $n\geq 3$, 
\begin{align}
\nonumber
\Pr\left(\bigcap_i\overline{A}_{iji'j'}\right)&\geq (1-2/n(n-1))^{\left(\frac{n^2}{k}\right){k\choose 2}}\\
\nonumber
-\log \Pr\left(\bigcap_i\overline{A}_{iji'j'}\right) &< \left(\frac{n^2}{k}\right){k\choose 2}\log(1-2/n(n-1))\\
\nonumber
&<2\left(\frac{2n^2}{kn(n-1)}\right){k \choose 2}\\
\nonumber
&<\left(\frac{8}{k}\right){k\choose 2}\\
\label{eq:latin}
&\leq 4(k-1).
\end{align}
Let $D\subset\FS$ be all encodings of permutations of $A$ that are Latin Transversals. $\K(D|A)=O(1)$. We recall that $P$ is the uniform distribution over all permutation of $A$. By the Equation \ref{eq:latin}, $-\log P(D)<4(k-1)$. Thus by Theorem \ref{thr:el} and Lemma \ref{lmm:consH}, for  $n\geq 3$, there exists a permutation $\tau\in D$ that is a Latin Transversal and has complexity
$$\K(\tau)\lel \K(P) - \log P(D) +\I(D;\ch)\lel \K(n) +4(k-1) + \I(A;\ch).$$
\qed
\end{prf}

\subsection{{\sc Function-Minimization}}
Given computable functions $\{f_i\}_{i=1}^n$, where each $f_i:\N\rightarrow\N\cup\infty$, the goal of {\sc Function-Minimization} is to find numbers $\{x_i\}_{i=1}^n$, that minimizes $\sum_{i=1}^nf_i(x_i)$. Let $p:\N\rightarrow\R_{\geq 0}$ be a computable probability measure where $\E_p[f_i]\in\R$ for all $i=1,\dots,n$. We define 
a computable  probability $P:\FS\rightarrow\R_{\geq 0}$ where $P(\langle a_1\rangle\langle a_2\rangle\dots\langle a_n\rangle) = \prod_{i=1}^np(a_i)$. $\K(P)\lea \K(p,n)$. Let $D'$ be a (potentially infinite) set of strings where $x\in D$ iff $x=\langle a_1\rangle\langle a_2\rangle\dots\langle a_n\rangle$ and
$$\sum_{i=1}^nf_i(a_i) \leq \left\lceil2\sum_{\{b_i\}}\left(\prod_{i=1}^np(b_i)\right)\sum_{i=1}^nf_i(b_i)\right\rceil=\left\lceil2\sum_{i=1}^n\E_{p}[f_i]\right\rceil.$$ 
Let $\tau=\ceil{2\sum_{i=1}^n\E_{p}[f_i]}$. By the Markov inequality, let the finite set $D\subseteq D'$ be constructed from $( p,\{f_i\},\tau)$, such that $P(D) >1/2$ and $\K(D|( p,\{f_i\},\tau))=O(1)$. By Theorem \ref{thr:el} and Lemma \ref{lmm:consH}, there a string $x\in D$
such that 
\begin{align*}
\K(x)&\lel -\log \m(D)+\I(D;\ch)\\
&\lel \K(P) -\log P(D)+\I(( p,\{f_i\},\tau);\chi)\\
&\lel \K(p,n)+\I(( p,\{f_i\},\tau);\chi).
\end{align*}

Thus given any computable probability $p$ and functions $\{f_i\}_{i=1}^n$, there are numbers $\{x_i\}_{i=1}^n$ such that $\sum_{i=1}^nf(x_i)\leq \ceil{2\sum_{i=1}^n\E_p[f_i]}=\tau$ and $\K(\{x_i\}_{i=1}^n)\lel \K(n,P)+\I(( p,\{f_i\},\tau);\ch)$. Note that there is a version of these results when the   functions are uncomputable, but this is out of the scope of the paper.

An instance of this formulation is as follows. Let $n=1$ and $f_1(a) = [a>2^m]\infty + [a\leq 2^m]2^{m-\K(a|m)}$. Let $p(a)=[a\leq 2^m]2^{-m}$. Thus this example proves there exists a number $x$ such that $f_1(x)\leq \ceil{2\E_p[f_1]}\leq 2$. Furthermore $$\K(x)\lel\K(p)+\I(( p,f_1);\ch)\lel \K(m)+\I(( m,f_1);\ch).$$ But if $f_1(x)\leq 2$, by the definition of $f_1$, this means $\K(x)\geq m-1$. This means $m\lel\I(( m,f_1);\ch)\lel \I(f_1;\ch)$. This makes sense because $f_1$ is a deficiency of randomness function and therefore $m\lel \K(f_1)$ and $\K(f_1|\ch)\lea\K(m)$.
\subsection{{\sc Super-Set}}
Given a finite set $S\subseteq \BT^n$, the goal of {\sc Super-Set} is to find a set $T\supseteq S$, $T\subseteq\BT^n$ that minimizes $|T|$. 
\begin{thr}
Given $m\leq n$, $S\subseteq \BT^n$, $|S|<2^{n-m-1}$ there exists a $T\supseteq  S$, $T\subseteq\BT^n$ $|T| =2^{n-m}$, $\K(T) \lel \K(n,m) + (m+1)|S|+ \I(( S,m);\ch)$.
\end{thr}
\begin{prf}
Let $P:\FS\rightarrow \R_{\geq 0}$ be the the uniform distribution over all sequences of size $2^n$ that have exactly $2^{n-m}$ 1s. Let $D\subset\BT^{2^n}$ consist of all sequences  $x_R\in \BT^{2^n}$ that encode sets $R\subseteq \BT^n$ in the natural way such that $R\supseteq S$ and $|R|=2^{m-n}$. Thus if $x\in D$ then $x$ has $2^{n-m}$ 1s. $P(D) =$
$$
\left(\frac{2^{n-m}}{2^n}\right)\left(\frac{2^{n-m-1}}{2^n-1}\right)\dots\left(\frac{2^{n-m}-|S|}{2^n-|S|}\right)\geq\left(\frac{2^{n-m}-|S|}{2^n-|S|}\right)^{|S|}\geq \left(\frac{2^{n-m-1}}{2^n}\right)^{|S|}=2^{-(m+1)|S|}.
$$
$\K(D|( S,m))=O(1)$. Thus by Theorem \ref{thr:el} and Lemma \ref{lmm:consH}, there exists a $t\in D$, such that 
$\K(t)\lel \K(P) - \log P(D) + \I(D;\ch)\lel \K(n,m) +(m+1)|S|+\I(( S,m);\ch)$. This $t$ encodes a set $T\supseteq S$, $T\subseteq\BT^n$ such that  $|T|=2^{n-m}$.

\qed

\end{prf}
  \subsection{{\sc Even-Odds}}
\label{sec:EvenOdds}

We define the following win/no-halt game, entitled {\sc Even-Odds}. There are $N$ rounds. At round $1$, the environment $\q$ secretly records bit $e_1\in\{0,1\}$. It sends an empty message to the agent who responds with bit $a_1\in\{0,1\}$. The agent gets a point if $e_1\oplus b_1=1$. Otherwise the agent loses a point. For round $i$, the environment secretly selects a bit $e_i$ that is a function of the previous agent's actions $\{a_j\}_{j=1}^{i-1}$ and sends an empty message to the agent, which responds with $a_i$ and the agent gets a point if $e_i\oplus a_i=1$, otherwise it loses a point. The agent wins after $N$ rounds if it has a score of at least $\sqrt{N}$.

\begin{thr}
\label{thr:EvenOdds}
For large enough $N$, there is a deterministic agent $\p$ that can win {\sc Even-Odds} with $N$ rounds, with complexity $\K(\p)\lel \I(\q;\ch)$.
\end{thr}
\begin{prf}
We describe a probabilistic agent $\p'$. At round $i$, $\p'$ submits 0 with probability 1/2. Otherwise it submits 1. By the central limit theorem, for large enough $N$, the score of the probabilistic agent divided by $\sqrt{N}$ is $S\sim\mathcal{N}(0,1)$. Let $\Phi(x)>\Pr[S>x]$. A common bound for $\Phi(x)$ is
\begin{align*}
\Phi(x)&>\frac{1}{2\pi}\frac{x}{x^2+1}e^{-x^2/2}\\
\Phi(1)&>\frac{1}{4\pi}e^{-1/2}>\frac{1}{8\pi}.
\end{align*}
Thus when $S\geq 1$, the score is at least $\sqrt{N}$. Thus $\p'$ wins with probability at least $p=\frac{1}{8\pi}$. Thus by Theorem \ref{thr:probwin}, there exists a deterministic agent $\p$ that can beat $\q$ with complexity
\begin{align*}
\K(\p)&\lel \K(\p')-\log p+\I(( p,\p',\q);\ch)\lel \I(\q;\ch).
\end{align*}
\qed
\end{prf}

\subsection{{\sc Graph-Navigation}}

The win/no-halt game is as follows. The environment $\q$ consists of $(G,s,r)$. $G=(E,V)$ is a non-bipartite graph with undirected edges, $s \in V$ is the starting vertex, and $r \in V$ is the goal vertex. 
Let $t_G$ be the time it takes for any random walk starting anywhere to converge to the stationary distribution $\pi(v)$, for all $v \in V$, up to a factor of 2.

There are $t_G$ rounds and the agent starts at $s\in V$. At round 1, the environment gives the agent the degree $s\in V$, $\mathrm{Deg}(s)$. The agent picks an number between 1 and $\mathrm{Deg}(s)$ and sends it to $\q$. The agent moves along the edge the number is mapped to and is given the degree of the next vertex it is on. Each round's mapping of numbers to edges to be a function of the current vertex, round number, and the agent's past actions. This process is repeated $t_G$ times. The agent wins if it is on $r\in V$ at the end of round $t_G$.

\begin{thr}
There is a deterministic agent $\p$ that can win the {\sc Graph-Navigation} game with complexity $\K(\p)\lel \log |E|+\I((G,s,r);\ch)$.
\end{thr}
\begin{prf}
 It is well known (see \cite{Lovasz96}), if $G$ is non-bipartite,  a random walk starting from any vertex will converge to a stationary distribution $\pi(v) = \mathrm{deg}(v)/2|E|$, for each $v \in V$. 

A probabilistic agent $\p'$ is defined as selecting each edge with equal probability. After $t_G$ rounds, the probability that $\p'$ is on the goal $r$ is close to the stationary distribution $\pi$. More specifically the probability is $\gem 1/|E|$. Thus by Theorem \ref{thr:probwin}, there is a deterministic agent $\p$ that can find $r$ in $t_G$ turns and has complexity $\K(\p')\lel \log |E| + \I((G,s,t);\ch)$. 

\end{prf}
\subsection{{\sc Penalty-Tests}}
An example penalty game is as follows. The environment $\q$ plays a game for $N$ rounds, for some very large $N\in\N$, with each round starting with an action by $\q$. At round $i$, the environment gives, to the agent, a program to compute a probability $P_i$ over $\N$. The choice of $P_i$ can be a computable function of $i$ and the agent's previous turns. The agent responds with a number $a_i\in\N$. The environment gives the agent a penalty of size $T_i(a_i)$, where $T_i:\N\rightarrow\Q_{\geq 0}$ is a computable test, with $\sum_{a\in\N}P_i(a)T_i(a)<1$. After $N$ rounds, $\q$ halts.
\begin{thr}
There is a deterministic agent $\p$ that can receive a penalty $<2N$ and has complexity $\K(\p)\lel \I(\q;\ch)$.
\end{thr}
\begin{prf}

A very successful probabilistic agent $\p'$ can be defined. Its algorithm is simple. On receipt of a program to compute $P_i$, the agent randomly samples a number $\N$ according to $P_i$. At each round the expected penalty is $\sum_{a}P_i(a)T_i(a)<1$, so the expected penalty of $\p$ for the entire game is $<N$. Thus by Corollary \ref{cor:pengame}, there is a deterministic agent $\p$ such that
\begin{enumerate}
\item $\p$ receives a penalty of $<2N$,
\item  $\K(\p)\lel \I(\q;\ch)$.
\end{enumerate}\qed
\end{prf}

Let $\q$ be defined so that $P_i(a) = [a\leq 2^i]2^{-i}$ and $T_i = [a\leq 2^i]2^{i-\K(a|i)}$. Thus each $T_i$ is a randomness deficiency function. The probabilistic algorithm $\p'$ will receive an expected penalty $<N$. However any deterministic agent $\p$ that receives a penalty $<2N$ must be very complex, as it must select many numbers with low randomness deficiency. Thus, by the bounds above, $\I(\q;\ch)$ must be very high. This makes sense because $\q$ encodes $N$ randomness deficiency functions.

\subsection{{\sc Cover-Time}}
We define the following interactive penalty game. Let $G = (E,V)$ be a graph consisting of $n$ vertices $V$ and undirected edges $E$. 
The environment $\q$ consists of $(G,s,\ell)$. $G=(E,V)$ is a non-bipartite graph with undirected edges, $s \in V$ is the starting vertex. $\ell$ is a mapping from numbers to edges to be described later.

The agent starts at $s\in V$. At round 1, the environment gives the agent the degree $s\in V$, $\mathrm{Deg}(s)$. The agent picks a number between 1 and $\mathrm{Deg}(s)$ and sends it to $\q$. The agent moves along the edge the number is mapped to and is given the degree of the next vertex it is on. Each round's mapping of numbers to edges, $\ell$, is a computable function of the current vertex, round number, and the agent's past actions.
The game stops if the agent has visited all vertices and the penalty is the number of turns the agents takes.
\begin{thr}
There is a deterministic agent $\p$ that can play against {\sc Cover-Time} instance $(G,S,\ell)$, $|G|=n$, and achieve penalty  $\frac{8}{27}n^3+o(n^3)$ and $\K(\p)\lel \I((G,s,\ell);\ch)$.
\end{thr}
\begin{prf}
A probabilistic agent $\p'$ is defined as selecting each edge with equal probability. Thus the agent performs a random walk. The game halts with probability 1. Due to \cite{Feige95}, the expected time (i.e. expected penalty) it takes to reach all vertices is $\frac{4}{27}n^3+o(n^3)$. Thus by Corollary \ref{cor:pengame} there is a deterministic agent $\p$ that can reach each vertex with a penalty of $\frac{8}{27}n^3+o(n^3)$ and has complexity
 $$\K(\p)\lel \K(\p')+\I((G,s,\ell);\ch)\lel\I((G,s,\ell);\ch).$$ \qed
\end{prf}

\subsection{{\sc Min-Cut}}
\begin{figure}     
  \centering
  \includegraphics[width=6cm]{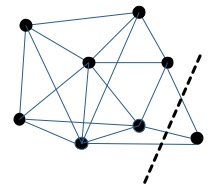}
  \caption{A graphical depiction of a minimum cut. By removing the edges along the dotted line, two components are created.}
  \label{fig:cut}
\end{figure}

We define the following win/no-halt game, entitled {\sc Min-Cut}. The game is defined by an undirected graph $G$ and a mapping $\ell$ from numbers to edges. At round $i$, the environment $\q$ sends the number of edges of $G$. The player responds with a number. The environment maps the number to an edge, and this mapping can be a function of the round number and player's previous actions. The environment then contracts the graph $G$ along the edge. The game halts when the graph $G$ has contracted into two vertices. The player wins if the cut represented by the contractions is a min cut. A minimum cut of a graph is the minimum number of edges, that when removed from the graph, produces two components. A graphical depiction of a min cut can be seen in Figure \ref{fig:cut}.
\begin{thr}
There is a deterministic agent $\p$ that can play against {\sc Cover-Time} instance $(G,S,\ell)$, $|G|=n$, such that $\K(\p)\lel 2\log n + \I((G,\ell);\ch)$.
\end{thr}
\begin{prf}
We define the following randomized agent $\p'$. At each round, $\p'$ chooses an edge at random. Thus the interactions of $\p'$ and $\q$ represent an implementation of Karger's algorithm. Karger's algorithm has an $\Omega(1/n^2)$ probability of returning a min-cut. Thus $\p'$ has an $\Omega(1/n^2)$ chance of winning. By Theorem \ref{thr:probwin}, there exist a deterministic agent $\p$ and $c$ where $\p$ can beat $\q$ and has complexity $\K(\p)\lel\K(p')-\log c/n^2+\I( \q;\ch)\lel 2\log n + \I((G,\ell);\ch)$.\qed
\end{prf}
 \subsection{{\sc Vertex-Transitive-Graph}}
 
\begin{figure}     
  \centering
  \includegraphics[width=6cm]{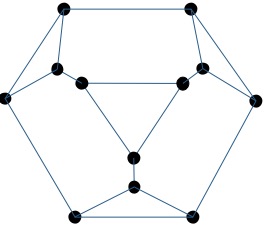}
  \caption{An example vertex-transitive graph.}
  \label{fig:transitive}
\end{figure}

 We describe the following graph based game. The environment $\q=(G,\ell,u,2k)$ consists of an undirected vertex-transitive graph $G=(V,E)$, a start vertex $u\in V$, the number of rounds $2k$, and a mapping $\ell$ from numbers to vertices. A vertex-transitive graph $G=(V,E)$ has the property that for any vertices $u,v\in V$, there is an automorphism of $G$ that maps $u$ into $v$. An example of a vertex transitive graph can be seen in Figure \ref{fig:transitive}. At round 1, the agent starts at vertex $u\in V$ and the environment send to the agent the degree of $u$. The agent picks a number from 1 to $\mathrm{Deg}(u)$ and the environment moves the agent along the edge specified by the mapping $\ell$ from numbers to edges. The mapping $\ell$ can be a function of the round number and the agents previous actions. The agent wins if after $2k$ rounds, the agent is back at $u$.
 \begin{thr}
 There is a deterministic agent $\p$ that can win at the {\sc Vertex-Transitive-Graph} game $(G=(V,E),\ell,u,k)$, $|V|=n$ with complexity $\K(\p)\lel \log n + \I((G,\ell,u,k);\ch)$.
  \end{thr}
  \begin{prf}
  We define the following randomized agent $\p'$. At each round, after being given the degree $d$ of the current vertex, $\p'$ chooses a number randomly from 1 to $d$. $\K(\p')=O(1)$. This is equivalent to a random walk on $G$. Let $P^l(u,v)$ denote the probability that a random walk of length $l$ starting at $u$ ends at $v$. Then due to   
\cite{AlonSp04}, for vertex-transitive graph $G$,
$$P^{2k}(u,u)\geq P^{2k}(u,v).$$
So after $2k$ rounds the randomized agent is back at $u$ with probability $P^{2k}(u,u)\geq 1/n$, which lower bounds the winning probability of $\p'$ against $\q$. By Theorem \ref{thr:probwin}, there exists a deterministic agent $\p$ that can beat $\q$ with complexity
$$\K(\p)\lel\K(\p') + n + \I(( n,\p',\q);\ch)\lel n+\I((G,\ell,u,k);\ch).$$\qed
  \end{prf}

\subsection{{\sc Classification}}
In machine learning, {\sc Classification} is the task of learning a binary function $c$  from $\N$ to bits $\BT$. The learner is given a sample consisting of pairs $(x,b)$ for string $x$ and bit $b$ and outputs a binary classifier $h:\N\rightarrow\BT$ that should match $c$ as much as possible. Occam's razor says that ``the simplest explanation is usually the best one.'' Simple hypothesis are resilient against overfitting to the sample data. The question is, given a particular problem in machine learning, how simple can the hypotheses be?

We use a probabilistic model. The target concept is modeled by a random variable $\mathcal{X}$ with distribution $p$ over ordered lists of natural numbers. The random variable $\mathcal{Y}$ models the labels, and has a distribution over lists of bits, where the distribution of $\mathcal{X}\times\mathcal{Y}$ is $p(x,y)$ with conditional probability requirement $p(y|x)=\prod_{i=1..|x|}p(y_i|x_i)$. Each such $(x_i,y_i)$ is a labeled sample. A binary classifier $f$ is consistent with labelled samples $(x,y)$, if for all $i$, $f(x_i)=y_i$. Let $\Gamma(x,y)$ be the minimum Kolmogorov complexity of a classifier consistent with $(x,y)$. $\mathcal{H}(\mathcal{Y}|\mathcal{X})$ is the conditional entropy of $\mathcal{Y}$ given $\mathcal{X}$.\\

\begin{thr}$ $\\
\vspace*{-0.4cm}
\begin{enumerate}
	\item $\mathcal{H}(\mathcal{Y}|\mathcal{X})\leq\E[\Gamma(\mathcal{X},\mathcal{Y})]\lel \mathcal{H}(\mathcal{Y}|\mathcal{X})+\K(p)$.
\item \textit{For each $c,b\in\N$, there exists random labeled samples $\mathcal{X}\times \mathcal{Y}$ with distribution $p$, such that, up to precision $O(\log cb)$, $\E[\Gamma(\mathcal{X},\mathcal{Y})]=b+c$, $\mathcal{H}(\mathcal{Y}|\mathcal{X})=b$, and $\K(p)= c$.	 }
 \end{enumerate}
 \end{thr}
\begin{prf}
We start with the lower bound of part 1.	$\E[\Gamma(\mathcal{X},\mathcal{Y})]=\sum_xp(x)\sum_yp(y|x)\Gamma(x,y)$. Each $\Gamma(x,y)$ represents a self-delimiting program to compute a classifier $f$ such that $f(x_i)=y_i$. Thus if $y\neq y'$, $\Gamma(x,y)$ and $\Gamma(x,y')$ represents two programs $v$ and $v'$ such that $v\not\sqsubseteq v'$ and $v'\not\sqsubseteq v$. Thus for a fixed $x$, ranged over $y$,  $\Gamma(x,y)$
represents the length of a self-delimiting code. Due to properties of conditional entropy, which is minimal over all self-delimiting codes, 
$$\E[\Gamma(\mathcal{X},\mathcal{Y})]=\sum_xp(x)\sum_yp(y|x)\Gamma(x,y)\geq \sum_xp(x)\sum_yp(y|x)(-\log p(y|x))=\mathcal{H}(\mathcal{Y}|\mathcal{X}).$$

We now prove the upper bound of part 2. To do so, we need the following lemma.
The following lemma is perhaps surprising because it shows that the $\I(\cdot;\mathcal{H})$ terms in inequalities can be removed by averaging over a computable probability.
\begin{lmm}
	\label{lmm:prodh}
	For computable probability $p$, $\sum_xp(x)\I(x;\mathcal{H})\lea \K(p)$.
\end{lmm}
\begin{prf}
	This follows from Theorem 3.1.3 in \cite{Gacs21}, and we will reproduce its arguments. Since $\K(x/\mathcal{H})$ is the length of a self delimiting code,
	 $$\sum_xp(x)\K(x/\mathcal{H})\geq \mathcal{H}(p),$$
	 where $\mathcal{H}(p)$ is the entropy of $p$. Furthermore, for all $x\in\FS$, $\K(x)\lea-\log p(x)+\K(p)$. Therefore $$\sum_xp(x)\K(x)\lea \sum_xp(x)(-\log p(x))+\K(p)\lea \mathcal{H}(p)+\K(p).$$
	 So
	 \begin{align*}
	 \sum_xp(x)\I(x;\mathcal{H})
	 &= \sum_xp(x)\left(\K(x)-\K(x/\mathcal{H})\right)
	 \lea \mathcal{H}(p)+\K(p)-\sum_xp(x)\K(x/\mathcal{H})
	\lea \K(p).
	 \end{align*}\qed
\end{prf}$ $\\

Binary classifiers are identified by infinite sequences $\alpha\in\IS$. We define the computable measure $S:\FS\rightarrow\R_{\geq 0}$ over $\IS$, where $S(x)=\prod_{n=1..|x|}p(x_n|n)$, where $\K(S|p)=O(1)$. Let $\{(x_i,y_i)\}$ be a set of labelled samples and we define clopen set $C_{x,y}=\{\alpha:\alpha\in\IS,\alpha[x_i]=y_i\}$. Then $S(C_{x,y})=p(y|x)$. By Theorem \ref{thr:mon}, relativized to $p$,
	\begin{align*}
	\min_{\alpha\in C_{x,y}}\K(\alpha|p)&\lel\K(S|p) -\log S(C_{x,y})+\I(C_{x,y};\mathcal{H}|p)\\
		&\lel -\log S(C_{x,y})+\I(C_{x,y};\mathcal{H}|p)\\	
	&\lel -\log p(y|x)+\I(C_{x,y};\mathcal{H}|p)
	\end{align*}
Averaging over all $x$ and $y$ using probability $p$, one gets
\begin{align}
\label{eq:1}
\sum_{x,y}p(x,y) \min_{\alpha\in C_{x,y}}\K(\alpha|p)&\lel \sum_{x,y}p(x,y)(-\log p(y|x))+\sum_{x,y}p(x,y)\I(C_{x,y};\mathcal{H}|p).
	\end{align}
	Applying Lemma \ref{lmm:prodh} relative to $p$, we get
		\begin{align}
		\label{eq:2}
		\sum_{x,y}p(x,y)\I( C_{x,y};\mathcal{H}|p)=\sum_{x,y}p( C_{x,y})\I( C_{x,y};\mathcal{H}|p)&\lea \K(p|p)=O(1).
	\end{align}
	Combining equations \ref{eq:1} and \ref{eq:2},
	\begin{align*}
	\sum_{x,y}p(x,y) \min_{\alpha\in C_{x,y}}\K(\alpha|p)&\lel \sum_{x,y}p(x,y)(-\log p(y|x))\\
		\left(\sum_{x,y}p(x,y) \min_{\alpha\in C_{x,y}}\K(\alpha)\right)-\K(p)&\lel \sum_{x,y}p(x,y)(-\log p(y|x))\\
\E[\Gamma(\mathcal{X},\mathcal{Y})] &\lel \mathcal{H}(\mathcal{Y}|\mathcal{X})+\K(p).
	\end{align*}

We now prove part 2. We ignore all $O(\log cd)$ terms. So equality $=$ is equivalent to $=\pm O(\log cd)$. We define a probability $p(x,y)$ over the first $n=2c+2b+2$ numbers and corresponding bits. Thus we can describe $p$ as a probability measure over strings of size $n$, making sure to maintain $p$'s conditional probability restriction described earlier.
	
	Let $z\in\BT^c$ be a random string of size $c$, with $c\lea \K(z)$. For all strings $w\in\BT^b$ of size $b$, $p(\langle z\rangle\langle w\rangle)=2^{-b}$, with $\|\langle z\rangle\langle w\rangle\|=n$. $\mathcal{H}(\mathcal{Y}|\mathcal{X})=-\sum_{w\in\BT^b}2^{-b}(\log p(\langle z\rangle\langle w\rangle))=-\sum_{w\in\BT^b}2^{-b}(\log 2^{-b})=b$. Furthermore $\K(p)=c$. The infinite sequence  $\alpha=\langle z\rangle \langle w\rangle0^\infty$ realizes $\Gamma( \langle z\rangle\langle w\rangle )$ up to an additive constant for each $w\in\BT^b$. Thus $\K(\alpha)= \K(z,w)$.  
 $$\E[\Gamma(\mathcal{X},\mathcal{Y})]=2^{-b}\sum_{w\in\BT^b}\K(\langle z\rangle\langle w\rangle)=\K(z)+2^{-b}\sum_{w\in\BT^b}\K(w/z,\K(z)).$$ Using  Theorem 3.1.3 in \cite{Gacs21} conditioned on $\langle z,\K(z)\rangle$, we get that $\sum_{w\in\BT^b}2^{-b}\K(w/z,\K(z))=\mathcal{H}(\mathcal{U}_b)\pm \K(b/z,\K(z))=b$, where $\mathcal{U}_b$ is the uniform measure over strings of size $b$. So $\E[\Gamma(\mathcal{X},\mathcal{Y})]=\K(z)+b=b+c$.\qed
\end{prf}
%\bibliographystyle{alpha} 
%\bibliography{refnotes}

\end{document}